\documentclass[aps,prb,twocolumn,superscriptaddress,nofootinbib]{revtex4-2}

\usepackage[utf8]{inputenc}
\usepackage[T1]{fontenc}
\usepackage{graphicx}
\usepackage{amsmath}
\usepackage{amssymb}
\usepackage{bm}
\usepackage{textcomp}
\usepackage[colorlinks=true,linkcolor=blue,citecolor=blue,urlcolor=blue]{hyperref}
\usepackage{float}
\usepackage{placeins}

\graphicspath{{figures/}}

\newcommand{\MoSSe}{MoS$_2$/WSe$_2$}
\newcommand{\MoSMoSe}{MoS$_2$/MoSe$_2$}
\newcommand{\MoSt}{MoS$_2$}
\newcommand{\WSet}{WSe$_2$}
\newcommand{\MoSet}{MoSe$_2$}
\newcommand{\kvec}{\bm{k}}
\newcommand{\qvec}{\bm{q}}
\newcommand{\rootseven}{\mbox{$\sqrt{7}\times\sqrt{7}\,R21.787^{\circ}$}}
\newcommand{\fth}{f_{\mathrm{th}}}
\newcommand{\kt}{\kappa_{12}}
\newcommand{\kss}{\kappa_{22}}

\begin{document}

\title{Phononic Frequency Combs in Twisted Bilayer Transition Metal Dichalcogenide
       Heterostructures from First Principles}

\author{Naman Arora}
\affiliation{Department of Electrical and Electronics Engineering,
             Birla Institute of Technology and Science, Pilani,
             Dubai Campus, Dubai International Academic City, Dubai, UAE}
\author{Adarsh Ganesan}
\affiliation{Department of Electrical and Electronics Engineering,
             Birla Institute of Technology and Science, Pilani,
             Dubai Campus, Dubai International Academic City, Dubai, UAE}

\date{\today}

\begin{abstract}
Phononic frequency combs are equidistant spectral ladders generated when a driven
vibrational mode transfers energy to an autoparametrically excited subharmonic mode through
anharmonic intermode coupling. In previous studies, the coupling strength governing this energy transfer has typically been treated as a free parameter, extracted from a fabricated device, or adopted from an earlier calculation for a specific mode pair. As a result, whether a given crystal can support a phononic frequency comb, and the drive strength required to generate it, has not been predictable a priori. Here we
obtain the complete parameter set from density functional theory. Twisted bilayers of
\MoSSe{} and \MoSMoSe{} at the commensurate angle $21.787^{\circ}$ adopt space group $P3$
in a 42-atom \rootseven{} moir\'e supercell. The second-order force constants provide the mode
frequencies; the third-order constants provide the linewidths through the Boltzmann transport
equation and hence the quality factors; and the same third-order tensor projected onto the phonon eigenvectors of a chosen mode pair provides the anharmonic couplings $\kt$ and $\kss$.
Both bilayers constitute a mode pair within $0.8\%$ of the $2\!:\!1$ resonance with $\omega_1/\omega_2 = 1.9923$ in \MoSSe{} and $1.9943$ in \MoSMoSe{}. By integrating the
resulting two-mode equations, both modes generate combs through a discontinuous transition: at
$701\,\fth$ in \MoSSe{} the envelope variance rises from $2.1\times10^{3}$ to $1.37\times10^{4}$ and the tooth count of mode 1 from 7 to 16 within a single $5\,\fth$ increment,
identifying the onset as subcritical. The spectral entropy peaks \emph{before} this jump,
at $341\,\fth$. The two compounds carry comparable coupling, differing by
a factor of $1.2$. However, the comb excitation threshold is set by the quality-factors of signal modes: $Q_2 = 672$ for \MoSSe{} and  
$Q_2 = 3844$ for \MoSMoSe{}.
\end{abstract}

\maketitle

\section{Introduction}

A frequency comb is a spectrum of equidistant lines generated by nonlinear mixing, and its
optical realisation transformed frequency metrology by providing a coherent link between
optical and microwave frequencies~\cite{Cundiff2003,Udem2002,Kippenberg2011}. The
vibrational counterpart was demonstrated by Ganesan, Do and Seshia in an aluminium nitride-on-silicon micromechanical resonator~\cite{Ganesan2017}, where a driven mode at $\omega_1$ was found to
transfer energy to an autoparametrically excited mode near $\omega_1/2$, producing a ladder
of phase-coherent lines. The mechanism is distinct from the Kerr nonlinearity of optical
microcombs~\cite{DelHaye2007,Kippenberg2018}: it requires two modes in an approximate $2\!:\!1$ internal resonance and a
quadratic anharmonic coupling between them, so that the driven mode modulates the effective
spring constant of the subharmonic mode at twice its own resonance frequency. The comb spacing is not set
by a cavity free spectral range but by the slow-flow beat frequency, the residual detuning
between $\omega_1$ and $2\omega_2$~\cite{Ganesan2018}.

The conditions under which such a comb exists were established subsequently. Qi et al. derived the threshold drive from a slow-flow reduction of the two-mode
equations and showed that the threshold is set by the product of the two damping rates
divided by the coupling~\cite{Qi2020}. Combs have since been reported in quartz~\cite{Kubena2020},
carbon nanotubes~\cite{Nanotube2025}, lithium niobate~\cite{LiNbO2025}, capacitive CMOS-MEMS resonators~\cite{MEMS2024}, graphene membranes~\cite{Graphene2022} and ferroelectric hafnia-zirconia resonators~\cite{Gao2026}. Frequency combs have been generated in a single driven mode wherein the second mode is replaced by a negative nonlinear friction~\cite{Ochs2022}. Also, phononic combs have been observed to form spontaneously in the optical phonons of the van der
Waals crystals CrGeTe$_3$ and CrSiTe$_3$~\cite{Chen2025}. The nonlinear landscape
surrounding the comb is itself rich:
cyclic-fold bifurcations, torus doubling and coexisting regular and irregular solutions have
all been catalogued~\cite{Czaplewski2018,Strong2024}, and the sensitivity of comb observables to perturbation has been proposed as a sensing mechanism~\cite{Sensitivity2026}.
Combs have also been predicted in a bulk crystal: Rangwala and Ganesan proposed the idea of driving a Higgs-like phonon mode of hexagonal InMnO$_3$ to obtain combs through its coupling to the
Goldstone-like mode~\cite{Rangwala2026}. A distinct route to comb formation has been identified very recently. Using
molecular dynamics driven by a machine-learned force field trained on \textit{ab initio}
data, Jayakrishnan \textit{et al.}~\cite{Jayakrishnan2026} showed that a single isolated
phonon mode with strong self-anharmonicity and a few available scattering channels produces
a comb spontaneously, without any drive, through the coherent superposition of its
unequally spaced eigenstates. 

In this paper, we showcase phononic frequency comb generation in van der Waals heterostructures, formed by stacking two-dimensional layers with a controlled interlayer twist. These materials present themselves as interesting candidates for the realization of internal resonance~\cite{Geim2013,Novoselov2016}. The moir\'e superlattice folds the phonon dispersion of the constituent monolayers into a reduced Brillouin zone, multiplying the number of optical branches and thereby the number of mode pairs available to satisfy
$\omega_1 \approx 2\omega_2$; the formation of the moir'e superlattice introduces interlayer shear and layer-breathing modes that are absent in the constituent monolayers, with their frequencies evolving systematically with the interlayer twist.~\cite{Zhang2013,Lin2018,Quan2021,Lu2022,Maity2020}. In the twisted bilayer, four-phonon processes become the dominant decay channel, in contrast to the three-phonon processes that dominate in the monolayer. This pronounced enhancement of anharmonicity has been attributed directly to the interlayer coupling introduced by the twist~\cite{Alencar2025}.

Previously, Zalalutdinov
\textit{et al.} demonstrated frequency comb generation in acoustic cavities formed by \MoSt{}/h-BN heterostructures operating between 50 and 600~GHz with $f\times Q$ product in the order of $10^{14}$~\cite{Cui2021}. A molecular dynamics and
continuum study of twisted bilayer \MoSt{} nanomechanical resonators also showcased that the twist induces a $2\!:\!1$ internal resonance through mode splitting from the anisotropic interlayer
potential, and reported combs with their tooth spacing tunable by drive
amplitude~\cite{JAP2025}. While this work establishes phononic frequency-comb generation in a twisted bilayer, the nonlinear coupling coefficients are not derived directly from the underlying electronic structure; instead, the classical interatomic potential employed in the analysis is fitted to available data rather than derived from first-principles calculations.

In this paper, we compute the complete parameter set for two twisted bilayer
heterostructures, \MoSSe{} and \MoSMoSe{}, at the commensurate angle $21.787^{\circ}$. The mode frequencies follow from the
harmonic force constants; the quality factors from the three-phonon linewidths obtained from
the third-order constants through the Boltzmann transport equation; and the coupling
constants $\kt$ and $\kss$ from the same third-order tensor projected onto the phonon eigenvectors. We then integrate the coupled two-mode equations for the extracted model parameters.

We chose these two materials to answer two simple questions. A frequency comb needs two vibrations whose frequencies are in a 2:1 ratio. First, does a twisted bilayer produce such a pair naturally? Second, does this depend on what the bilayer is made of, or does it happen for different materials alike? We address this question by examining two different chemistries at the same interlayer twist angle. Second, once the resonance is established, what determines the strength of the resulting comb? \MoSSe{} and \MoSMoSe{} share a \MoSt{} layer and the same twist angle, but differ in whether the second layer changes both the metal and the chalcogen or only the chalcogen. This comparison therefore isolates the role of chemical asymmetry from that of the moir'e geometry. We find that the internal resonance is generic to the twisted-bilayer architecture, whereas the strength of the nonlinear coupling-- and consequently the comb-- is strongly chemistry dependent.

\section{Computational Methods}

The workflow comprises four stages, applied identically to both \MoSSe{} and \MoSMoSe{}: (i) relaxation of the twisted bilayer at fixed cell parameters, followed by a self-consistent ground-state calculation; (ii) calculation of second-order force constants using the finite-displacement method; (iii) calculation of third-order force constants from randomly displaced atomic configurations, followed by evaluation of phonon linewidths and quality factors; and (iv) projection of the third-order force-constant tensor onto the selected mode pair to determine the anharmonic coupling constants entering the frequency-comb equations.

\subsection{Structures and relaxation}

Twisted bilayers were constructed at the commensurate angle $21.787^{\circ}$~\cite{Shallcross2010}, which
generates a \rootseven{} moir\'e supercell on the triangular lattice of the transition metal
dichalcogenide monolayer. Each cell contains 42 atoms: 14 in the metal sublattices and 28 in
the chalcogen sublattices, distributed as \MoSt{} over \WSet{} in one compound and \MoSt{}
over \MoSet{} in the other. Starting geometries were taken from the twisted bilayer \MoSt{}
structures of Klein \textit{et al.}~\cite{Klein2024}, with the second layer substituted by
\WSet{} or \MoSet{}. The upper layer is rotated about an axis through a metal site.
The cells are periodic along the stacking axis with $c \approx 13$~\AA{}; they retain
space group $P3$ (No.~143, point group $3$, $C_3$). Removing the force constants that
couple each bilayer to its periodic image along $c$ and re-imposing the acoustic sum rule
shifts $\omega_1$ and $\omega_2$ by $-0.53\%$ and $-0.40\%$ in \MoSSe{} and by $-0.14\%$
and $-0.19\%$ in \MoSMoSe{}, and leaves both pairs within $1.0\%$ of the $2\!:\!1$
condition, so the contact across the cell boundary has a minimal effect on the comb modes.

Density functional theory calculations were performed with VASP~\cite{Kresse1996} using
projector-augmented-wave potentials~\cite{Blochl1994} and the PBE parameterisation of the
generalised gradient approximation~\cite{Perdew1996}. The Mo\_sv potential (14 valence
electrons, including the $4s$ and $4p$ semicore shells), W (6 valence electrons), S and Se
(6 valence electrons each) were used. A plane-wave cutoff of $380$~eV was employed
throughout, and the Brillouin zone of the moir\'e cell was sampled with an $11\times11\times1$ $\Gamma$-centred $\kvec$-mesh; a single $k_z$ point
is adequate because the coupling across the cell boundary is weak (Sec.~II~A). Van der Waals interactions were accounted for using Grimme’s DFT-D3 correction~\cite{Grimme2010}, which is essential for describing interlayer binding and coupling in layered systems, where these interactions influence the mode splitting investigated here.

Ionic positions were relaxed at fixed cell shape and volume (\texttt{ISIF}~$=2$) to preserve the commensurate moiré geometry and twist angle. Allowing the cell to relax could alter the imposed supercell geometry and compromise the controlled comparison between the two heterostructures. Fixing the cell while relaxing the internal coordinates allows the interlayer separation and registry-dependent corrugation to equilibrate under the dispersion correction. At small twist angles, such internal relaxation can drive substantial moiré reconstruction, producing commensurate domains separated by domain walls~\cite{Weston2020}. 
Electronic self-consistency was converged to $10^{-7}$~eV and relaxation
continued until all residual forces fell below $0.01$~eV\,\AA$^{-1}$
(\texttt{EDIFFG}~$=-0.01$). A self-consistent calculation on the relaxed structure, at the
same cutoff and $\kvec$-mesh, then provided the ground state from which the electronic
density of states and all subsequent force calculations were started.

Spin–orbit coupling (SOC) was neglected in the calculations. Its effect on phonon frequencies in \MoSt{}-based dichalcogenides is reported to be below 1$\%$~\cite{MolinaSanchez2011}. Since comb generation depends on the frequency ratio $\omega_1/\omega_2$ rather than on the absolute frequencies, a common-mode frequency shift would largely cancel in this ratio. The third-order force constants, which govern the anharmonic intermode coupling, are also expected to be less sensitive to SOC than the electronic structure itself. The heavier \WSet{} layer represents the more stringent case; we return to this limitation in Sec.~\ref{sec:limits}.

\subsection{Harmonic force constants}

Second-order force constants were obtained by the finite-displacement method within VASP
(\texttt{IBRION}~$=6$, \texttt{NSW}~$=1$), with \texttt{LEPSILON}~$=$~\texttt{.TRUE.} so
that the Born effective charges and the dielectric tensor were computed in the same run.
The resulting force constants were read by phonopy~\cite{Togo2015} and stored as
\texttt{fc2.hdf5}. The moir\'e cell itself was used as the supercell
($1\times1\times1$).
Phonon frequencies, eigenvectors and the density of states were evaluated on a
$80\times80\times1$ $\Gamma$-centred $\qvec$-mesh. The mesh is restricted to the plane in
order to describe scattering within the bilayer; sampling $q_z$ would add channels between
periodic images, which are not part of the bilayer.

\subsection{Anharmonic force constants by random displacement}
\label{sec:rd}

Calculating third-order force constants presents a computational challenge. A systematic finite-displacement approach for these supercells would require 27,722 configurations, rendering it computationally prohibitive at the chosen level of theory. We therefore used random displacement
sampling~\cite{Togo2023,Phono3py2023,Zhou2014,Tadano2014,Eriksson2019}: ten supercell configurations per compound were
generated with phono3py, with all atoms displaced simultaneously by $0.03$~\AA{} in random
directions, and the forces on each configuration were computed with VASP at the same
settings as above. The third-order tensor was obtained by fitting these forces, with the
harmonic part fixed to \texttt{fc2.hdf5}, and written as \texttt{fc3.hdf5} for use by
phono3py in the linewidth calculation below.

Because the ten configurations provide fewer independent equations than there are independent tensor components, the fitting problem is underdetermined and requires regularisation. We minimise
\begin{equation}
  \mathcal{L}[\Phi^{(3)}]
  = \bigl\lVert \bm{F} - \Phi^{(2)}\bm{u} - \tfrac{1}{2}\Phi^{(3)}\bm{u}\bm{u} \bigr\rVert^{2}
  + \alpha \bigl\lVert \Phi^{(3)} \bigr\rVert^{2},
  \label{eq:ridge}
\end{equation}
where $\bm{F}$ are the calculated forces, $\bm{u}$ the displacements and $\alpha = 10^{-3}$
the regularisation parameter, solved in closed form through the normal equations (ridge
regression~\cite{Hoerl1970}). The acoustic sum rule
\begin{equation}
  \sum_{j} \Phi^{(3)}_{\alpha\beta\gamma}(i,j,k) = 0
  \label{eq:asr}
\end{equation}
was then enforced by iterative redistribution of the residual onto the diagonal term,
together with the permutation symmetry
$\Phi^{(3)}_{\alpha\beta\gamma}(i,j,k) = \Phi^{(3)}_{\beta\alpha\gamma}(j,i,k)$, until the
maximum drift fell below $10^{-10}$~eV\,\AA$^{-3}$; convergence was reached in two
iterations. We explicitly acknowledge this limitation: the ten configurations provide only weak constraints on the third-order force-constant tensor, potentially leading to systematic underestimation of three-phonon scattering rates and, consequently, overestimation of the phonon quality factors. We quantify this in Sec.~\ref{sec:limits} and treat the $Q$ values as upper bounds throughout.

\subsection{Linewidths and quality factors}

Phonon linewidths were obtained from the imaginary part of the three-phonon self-energy
within the relaxation time approximation, as implemented in phono3py~\cite{Togo2023}. For
mode $\lambda$ at temperature $T$,
\begin{equation}
  \Gamma_{\lambda}(T)
  = \frac{18\pi}{\hbar^{2}} \sum_{\lambda'\lambda''}
    \bigl\lvert V^{(3)}_{\lambda\lambda'\lambda''} \bigr\rvert^{2}
    \, \delta\bigl(\omega_{\lambda} - \omega_{\lambda'} - \omega_{\lambda''}\bigr)
    \, \bigl[ n_{\lambda'} - n_{\lambda''} \bigr],
  \label{eq:gamma}
\end{equation}
with $n$ the Bose-Einstein occupations, evaluated at $T = 300$~K by tetrahedron integration~\cite{Blochl1994tet}
over the $80\times80\times1$ mesh. The quality factor follows as
\begin{equation}
  Q_{\lambda} = \frac{\omega_{\lambda}}{2\Gamma_{\lambda}} .
  \label{eq:Q}
\end{equation}
The $\Gamma$ point was located by minimising $\lvert\qvec\rvert$ over the irreducible mesh
rather than by taking the first index, which is not in general the zone centre in the
phono3py ordering.

\subsection{Anharmonic coupling constants}
\label{sec:kappa}

The quantity entering the comb equations is not the real-space tensor but its projection
onto the normal modes of the pair in question. For modes $j_1,j_2,j_3$ at $\Gamma$ with
eigenvectors $e^{(j)}_{i\alpha}$ and atomic masses $m_i$ ($i$ atom index, $\alpha$ Cartesian
index),
\begin{equation}
  V^{(3)}_{j_1 j_2 j_3}
  = \sum_{ijk} \sum_{\alpha\beta\gamma}
        \frac{\Phi^{(3)}_{\alpha\beta\gamma}(i,j,k)}{\sqrt{m_i m_j m_k}}\,
    e^{(j_1)}_{i\alpha} e^{(j_2)}_{j\beta} e^{(j_3)}_{k\gamma},
  \label{eq:V3}
\end{equation}
is evaluated by successive tensor contraction rather than by explicit summation, which reduces
the cost from $\mathcal{O}(N^{3})$ to $\mathcal{O}(N^{2})$ in the atom count and makes the
scan over all $126^{2}$ mode pairs tractable, scaling by the zero-point amplitude of each
mode,
\begin{equation}
  \kappa_{j_1 j_2 j_3}
    = V^{(3)}_{j_1 j_2 j_3}
    \sqrt{\frac{\hbar^{3}}{8\,\omega_{j_1}\omega_{j_2}\omega_{j_3}}} ,
  \label{eq:kappa}
\end{equation}
which carries units of energy. The dimensionless constants entering the equations of motion
are obtained by dividing by the driven-mode phonon energy $\hbar\omega_1$. 

Both coupling coefficients originate from the same cubic anharmonic term. In the potential energy,
$V = \tfrac{1}{3!}\sum\Phi_{jkl}q_jq_kq_l$, the three permutations of the index combination $(1,2,2)$ yield
$V \supset \tfrac{1}{2}\Phi_{122}q_1q_2^{2}$. Consequently, the coefficient of $x_1x_2$ in Eq.~(\ref{eq:ode2}) is $\Phi_{122}$, whereas that of $x_2^2$ in Eq.~(\ref{eq:ode1}) is $\tfrac{1}{2}\Phi_{122}$. Thus, $\kss=\kt/2$ exactly, and a single computed matrix element determines both coupling coefficients. We report $\lvert\kappa\rvert$ throughout, since the phase of each phonon eigenvector is arbitrary up to a sign, making the sign of $\kappa$ convention-dependent. The observable quantities considered here, including the threshold in Eq.~(\ref{eq:fth}), the comb-tooth amplitudes, and the scattering rates in Eq.~(\ref{eq:gamma}), depend on $\lvert\kappa\rvert^2$ and are therefore independent of this sign convention.

Candidate pairs were identified by scanning all mode pairs for
$\lvert \omega_1 - 2\omega_2 \rvert < 0.05$~THz and $\lvert\kt\rvert > 10^{-5}$~meV, which
returns 60 pairs in \MoSSe{} and 48 in \MoSMoSe{}. Among these candidates, we selected the mode pair that combines a large dimensionless $\lvert\kt\rvert$ with small detuning. Detuning is defined throughout as $\lvert\omega_1-2\omega_2\rvert/\omega_2$.

\subsection{Comb dynamics}

With the parameters fixed, the dynamics follow the two-mode equations of
Ref.~\onlinecite{Ganesan2017}, written in the dimensionless time $\tau = \omega_1 t$ and
displacement $X_j = x_j/x_0$:
\begin{align}
  \ddot{X}_1 + \frac{1}{Q_1}\dot{X}_1 + X_1 + \kss X_2^{2}
    &= f_0 \cos(\Omega_D \tau), \label{eq:ode1}\\
  \ddot{X}_2 + \frac{\Omega_2}{Q_2}\dot{X}_2 + \Omega_2^{2} X_2 + \kt X_1 X_2
    &= 0, \label{eq:ode2}
\end{align}
with $\Omega_2 = \omega_2/\omega_1$. The drive was set half a linewidth above resonance, $\Omega_D = 1 + 1/(2Q_1)$. The reference length $x_0$ enters $\kappa$ and $f_0$ with the same power and therefore cancels in the ratio $f_0/\fth$, which is used to report all results. Accordingly, displacement amplitudes are expressed in arbitrary units. The term $\kt X_1 X_2$ in
Eq.~(\ref{eq:ode2}) is of Mathieu type: with $X_1 \sim A\cos\tau$ it modulates the effective
stiffness of the second mode at $\omega_1 \approx 2\omega_2$, which is the condition for
parametric amplification. The term $\kss X_2^{2}$ in Eq.~(\ref{eq:ode1}) returns energy to
the driven mode, and the cascade of these two processes populates the comb.

The threshold above which the trivial solution $X_2 = 0$ loses stability is~\cite{Nayfeh1979,Qi2020}
\begin{equation}
  \fth = \frac{8\,\delta_1 \delta_2 \Omega_2}{\kt},
  \qquad \delta_1 = \frac{1}{2Q_1},\quad \delta_2 = \frac{\Omega_2}{2Q_2} .
  \label{eq:fth}
\end{equation}
Equations~(\ref{eq:ode1})--(\ref{eq:ode2}) were integrated with an explicit Runge-Kutta
scheme of order 4(5) with relative and absolute tolerances $10^{-9}$ and $10^{-12}$ and a
maximum step of $0.5$ in $\tau$, over $10^{4}$ cycles of $\omega_1$ sampled at
$5\times10^{6}$ points, from initial conditions
$X_1 = 1$, $X_2 = 0.01$; the small seed in the second mode plays the role of the thermal
fluctuation that initiates parametric growth in a physical system. Spectra were computed
from the final $15\%$ of the trajectory to exclude the transient. The drive was then swept
from $1$ to $1.01\times10^{4}\,\fth$ in \MoSSe{} and to $1.16\times10^{4}\,\fth$ in \MoSMoSe{}, in steps
of $5\,\fth$, with a full integration at each step.

Three diagnostics are tracked as the drive is swept. The comb tooth count is defined as the number of spectral peaks within $130$~dB of the maximum, with adjacent peaks separated by at least 30 frequency bins. Peaks are counted within a $\pm1.5$~THz window centred on $\omega_1$ for mode 1 and on $\omega_2$ for mode 2. The detection parameters are held fixed across all drive values to ensure comparability. The envelope variance is the variance of $\lvert\mathcal{H}[X_1]\rvert$, where $\mathcal{H}[X_1]$ denotes the Hilbert envelope of the driven mode. It quantifies the depth of amplitude modulation and, from a reservoir-computing perspective, characterises the temporal richness of the state that may support memory. The spectral entropy is defined as $H = -\sum_i p_i \ln p_i$~\cite{Powell1979}, where $p_i$ denotes the normalised power spectrum of $X_1$ after components below $4\%$ of the maximum are discarded. It quantifies spectral complexity: entropy is low for both a single tone and a regular comb, but higher for an irregular spectrum.
\section{Results}

\subsection{Moir\'e supercells and electronic structure}
\label{sec:structure}

\begin{figure*}
  \centering
  \includegraphics[width=\linewidth]{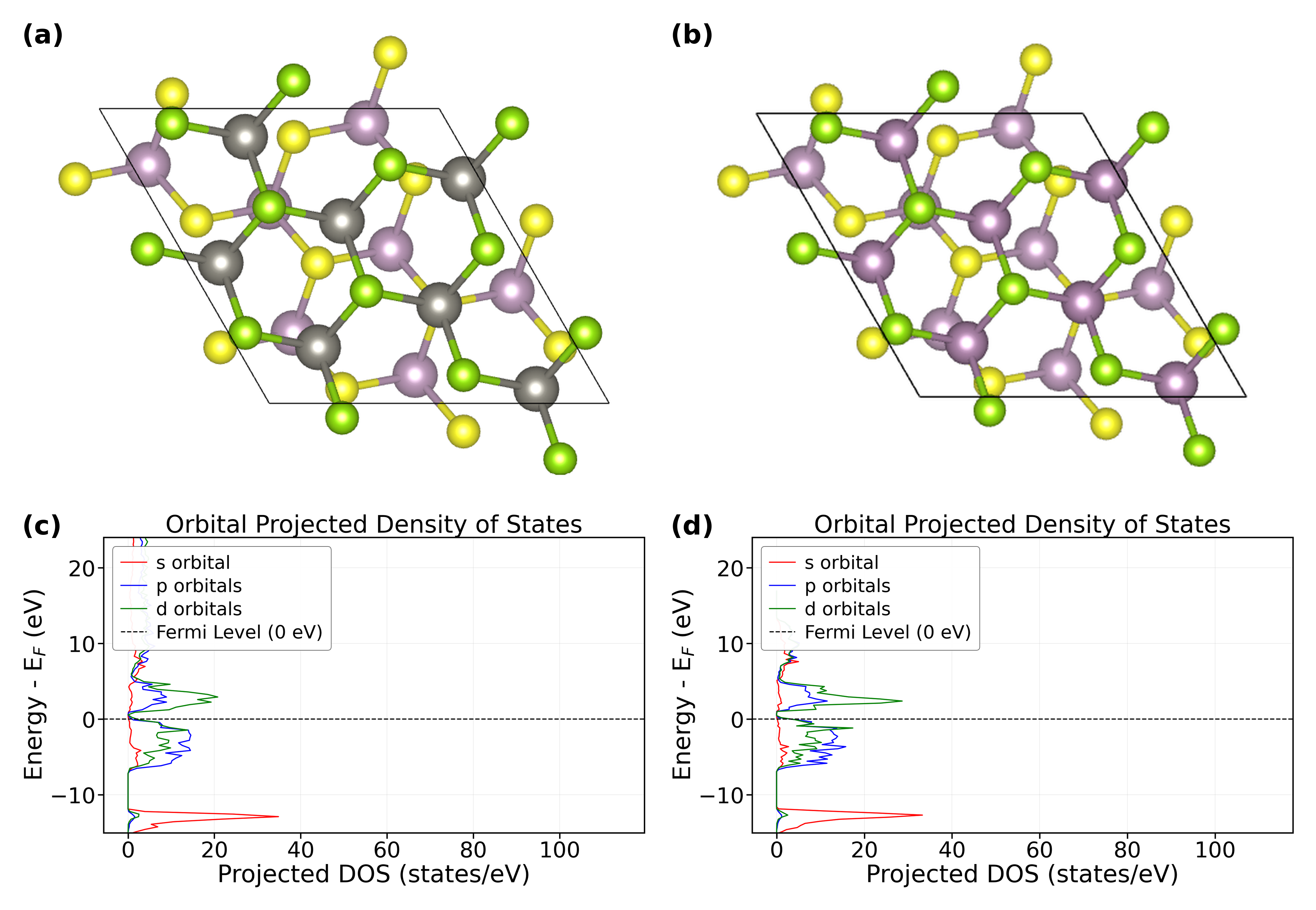}
  \caption{Relaxed \rootseven{} moir\'e supercells and their electronic structure. Left
           column, \MoSSe{}; right column, \MoSMoSe{}, both at the commensurate twist angle
           $21.787^{\circ}$. (a), (b) Top views along the stacking axis; each cell contains
           42 atoms and retains space group $P3$. Mo is shown in violet, W in grey, S in
           yellow and Se in green; in (b) both metal sublattices are Mo. The parallelogram
           marks the moir\'e cell. (c), (d) Orbital-projected ($s$, $p$, $d$) electronic
           density of states of the cells in (a) and (b), with the Fermi level at zero
           (dashed line). Both are semiconducting.}
  \label{fig:structure}
\end{figure*}

Figures~\ref{fig:structure}(a) and~\ref{fig:structure}(b) show the two relaxed cells viewed
along the stacking axis. The $21.787^{\circ}$ rotation carries one monolayer onto a
\rootseven{} supercell of the other, so that seven primitive cells of each layer are
contained in the moir\'e cell, outlined by the parallelogram, with the two lattices regaining registry at its corners. With three atoms per primitive cell in each monolayer, the moiré cell contains 21 atoms per layer and 42 atoms in total: 14 metal atoms and 28 chalcogen atoms. In Fig.~\ref{fig:structure}(a) the two metal sublattices are
distinguishable, the Mo sites of the \MoSt{} layer are in violet and the W sites of the \WSet{}
layer are in grey, and the rotation of one layer relative to the other is visible directly in
the misalignment of the metal--chalcogen bonds of the two layers. In
Fig.~\ref{fig:structure}(b) both metal sublattices are Mo, and the two layers are
distinguished only by their chalcogen, S in yellow and Se in green. Because the registry
between the layers varies across the cell, no two metal sites of opposite layers sit in the
same local stacking environment; this position-dependent interlayer registry is the moir\'e
modulation that the fixed-cell relaxation of Sec.~II~A is designed to preserve. The in-plane lattice constant of the supercell is $8.491$~\AA{} for \MoSSe{} and $8.357$~\AA{} for
\MoSMoSe{}, so the \WSet{} and \MoSet{} layers are held under in-plane compressive strain to
preserve the commensurate cell.

The rotation is taken about an axis through a metal site, which leaves that site invariant
and reduces the symmetry of the bilayer to $P3$ (No.~143, point group $3$, $C_3$). The threefold rotational symmetry is preserved, but inversion symmetry is absent, yielding a polar point group. This broken inversion symmetry lifts the parity-based selection rule that would otherwise forbid the cubic interaction term responsible for returning energy from the subharmonic to the driven mode. We revisit this mechanism in Sec.~\ref{sec:discussion}.

Figures~\ref{fig:structure}(c) and~\ref{fig:structure}(d) show the orbital-projected
electronic density of states of the same two relaxed cells, obtained from the
self-consistent calculation of Sec.~II~A. Both systems remain semiconducting after twisting, with a finite band gap at the Fermi level. Consequently, the phonon lifetimes calculated in Sec.~\ref{sec:pairs} are limited to phonon–phonon scattering, and electron–phonon scattering channels can be neglected. In both compounds, the states
bordering the gap are hybridised $d$ and $p$ states: the valence edge
carries comparable $p$ and $d$ weight, and the conduction edge is dominated by $d$ states.
The sharp $s$-projected peak deep in the valence band originates from the chalcogen $s$ states. The close similarity between the two spectra is consistent with their shared \MoSt{} layer and moiré geometry.

The two compounds share the \MoSt{} layer and twist angle, differing only in the composition of the second layer: both the metal and chalcogen are replaced in \MoSSe{}, whereas only the chalcogen is replaced in \MoSMoSe{}. The moiré geometry is therefore identical in both systems, allowing differences in comb behaviour to be attributed to the chemistry of the second layer rather than to changes in the superlattice geometry.

\subsection{Mode pairs and parametric resonance}
\label{sec:pairs}

The 42-atom cell carries 126 phonon branches extending to $\approx 14$~THz, and the scan of Sec.~\ref{sec:kappa} returns
60 pairs in \MoSSe{} and 48 in \MoSMoSe{} within the $0.05$~THz detuning window. The abundance of mode pairs satisfying the resonance condition is itself a key result, arising from the moiré geometry rather than the chemical composition. Folding the monolayer dispersion into the \rootseven{} Brillouin zone increases the number of phonon branches sevenfold. With 126 branches spanning $0.1$--$14.0011$~THz, the resulting high density of states makes near-resonant pairs satisfying $\omega_1 \approx 2\omega_2$ readily available without fine-tuning.

Table~\ref{tab:params} lists the selected pairs. In \MoSSe{} the pump mode lies at
$6.1619$~THz and the signal at $3.0929$~THz, giving $\omega_1/\omega_2 = 1.9923$ and a
detuning of $0.77\%$; in \MoSMoSe{} the corresponding frequencies are $11.6538$ and
$5.8436$~THz, with ratio $1.9943$ and detuning $0.57\%$.
\begin{table}[tbp]
  \caption{First-principles parameters of the selected mode pairs at $300$~K.}
  \label{tab:params}
  \begin{ruledtabular}
  \setlength{\tabcolsep}{4pt}
  \begin{tabular}{lcc}
    Quantity & \MoSSe{} & \MoSMoSe{} \\
    \colrule
    Atoms per cell            & 42      & 42      \\
    Space group               & $P3$      & $P3$      \\
    Branches                  & 126     & 126     \\
    Resonant pairs found      & 60      & 48      \\
    $\omega_1$ (THz)          & 6.1619  & 11.6538 \\
    $\omega_2$ (THz)          & 3.0929  & 5.8436  \\
    $\omega_1/\omega_2$       & 1.9923  & 1.9943  \\
    Detuning (\%)             & 0.77    & 0.57    \\
    $Q_1$                     & 5087    & 1825    \\
    $Q_2$                     & 672     & 3844    \\
    $\lvert\kt\rvert$ (meV)   & 0.0026  & 0.0040  \\
    $\lvert\kss\rvert$ (meV)  & 0.0013  & 0.0020  \\
    $\hbar\omega_1$ (meV)     & 25.48   & 48.20   \\
    $\lvert\kt\rvert$ (dimensionless)  & $1.00\times10^{-4}$ & $8.37\times10^{-5}$ \\
    $\lvert\kss\rvert$ (dimensionless) & $5.01\times10^{-5}$ & $4.18\times10^{-5}$ \\
  \end{tabular}
  \end{ruledtabular}
\end{table}

In \MoSSe{} the driven mode is more long-lived with $Q_1 = 5087$ and $Q_2 = 672$; however, in
\MoSMoSe{}, the signal mode is more long-lived with $Q_1 = 1825$ and $Q_2 = 3844$. Since the threshold
of Eq.~(\ref{eq:fth}) depends on the product $\delta_1\delta_2$, the two arrangements are not
equivalent: energy fed into a short-lived signal mode must be replenished continuously,
whereas a long-lived signal mode retains it. With $Q_2 = 672$ in \MoSSe{} against $Q_2=3844$ in
\MoSMoSe{}, the damping of the signal mode differs by a factor of $5.7$,
$\delta_2 = 3.73\times10^{-4}$ against $6.52\times10^{-5}$.

The coupling strengths, by contrast, are similar in the two compounds. The dimensionless coupling is $\kt = 1.00\times10^{-4}$ for \MoSSe{} and $8.37\times10^{-5}$ for \MoSMoSe{}, a ratio of 1.20. Evaluating Eq.~(\ref{eq:fth}) using the corresponding parameters yields $\fth = 1.47\times10^{-3}$ for \MoSSe{} and $8.56\times10^{-4}$ for \MoSMoSe{}. Thus, despite its slightly weaker coupling, \MoSMoSe{} has the lower absolute threshold, indicating that differences in damping outweigh the modest difference in coupling strength.

\subsection{Comb spectra}
\label{sec:spectra}

\begin{figure*}
  \centering
  \includegraphics[width=\linewidth]{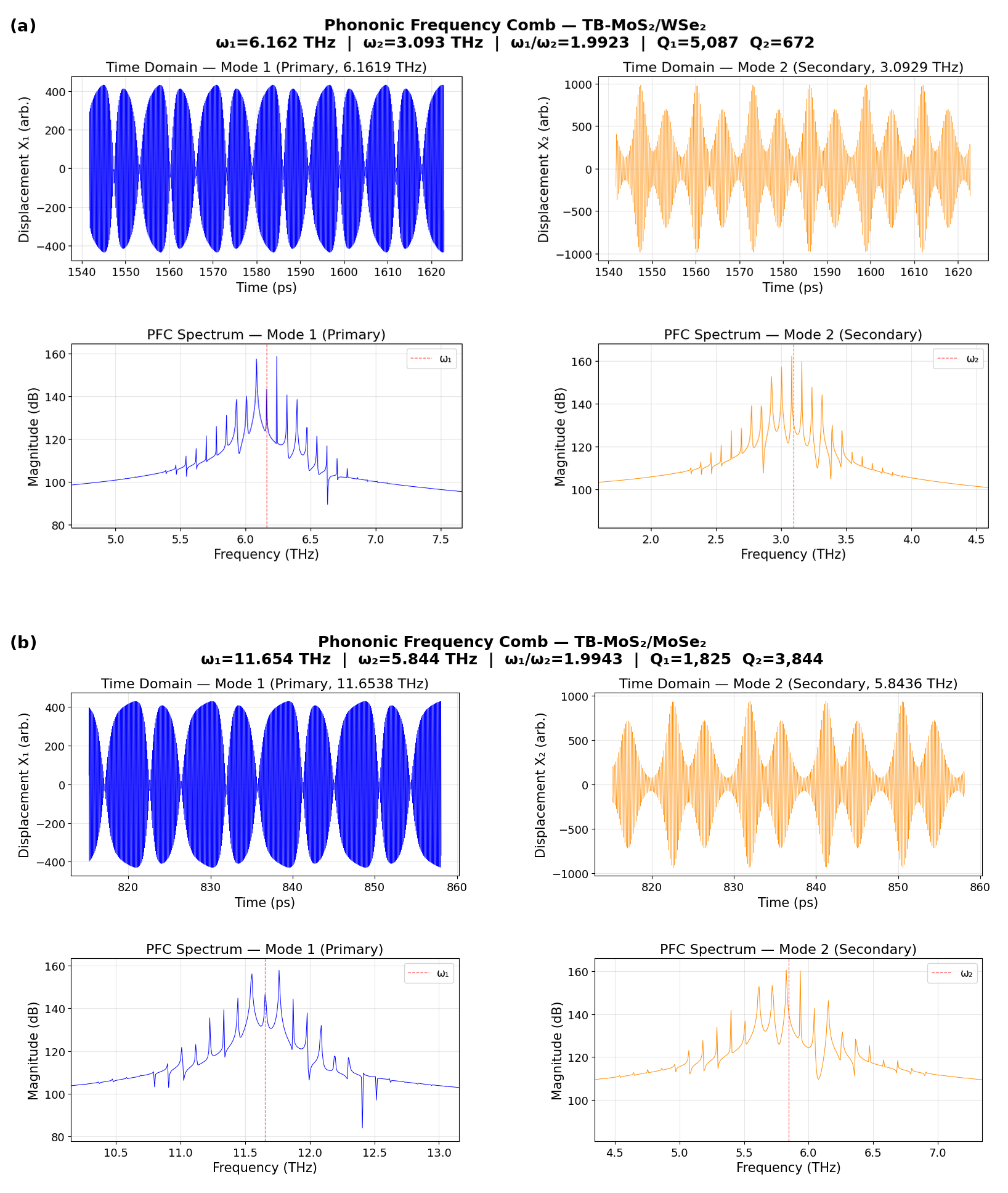}
    \caption{Phononic frequency combs in the two heterostructures at the same drive, $f_0 = 1186\,\fth$. (a) \MoSSe{}; (b) \MoSMoSe{}. In each, upper panels show the
           time-domain displacement of the driven mode (blue) and of the parametrically
           excited mode (orange); lower panels show the corresponding spectra about
           $\omega_1$ and $\omega_2$, with the mode frequencies marked by dashed lines. The
           envelope modulation in the time domain is the beating of the comb teeth against
           one another.}
  \label{fig:spectra}
\end{figure*}

Figure~\ref{fig:spectra} compares the two compounds at the same multiple of their
respective thresholds, $f_0 = 1186\,\fth$, \MoSSe{} in (a) and \MoSMoSe{} in (b). In both, the spectrum of the driven mode carries a set of
equidistant lines about $\omega_1$, and the signal mode carries the matching ladder about
$\omega_2$. The time domain shows the envelope modulation characteristic of several teeth
beating against one another. The resolved spectra in the lower panels show the individual teeth. In \MoSSe{} the teeth
are spaced by $\approx 77$~GHz, and the same spacing appears about $\omega_1$ and about
$\omega_2$, as required for two ladders generated by the same slow-flow beat. The
envelope period in the time domain is the inverse of this spacing. \MoSMoSe{} follows the same behaviour with fewer teeth, and its teeth are more widely
spaced, $\approx 107$~GHz, as the resolved spectra show directly.

\subsection{Drive sweep and the character of the onset}
\label{sec:sweep}

\begin{figure*}
  \centering
  \includegraphics[width=\linewidth]{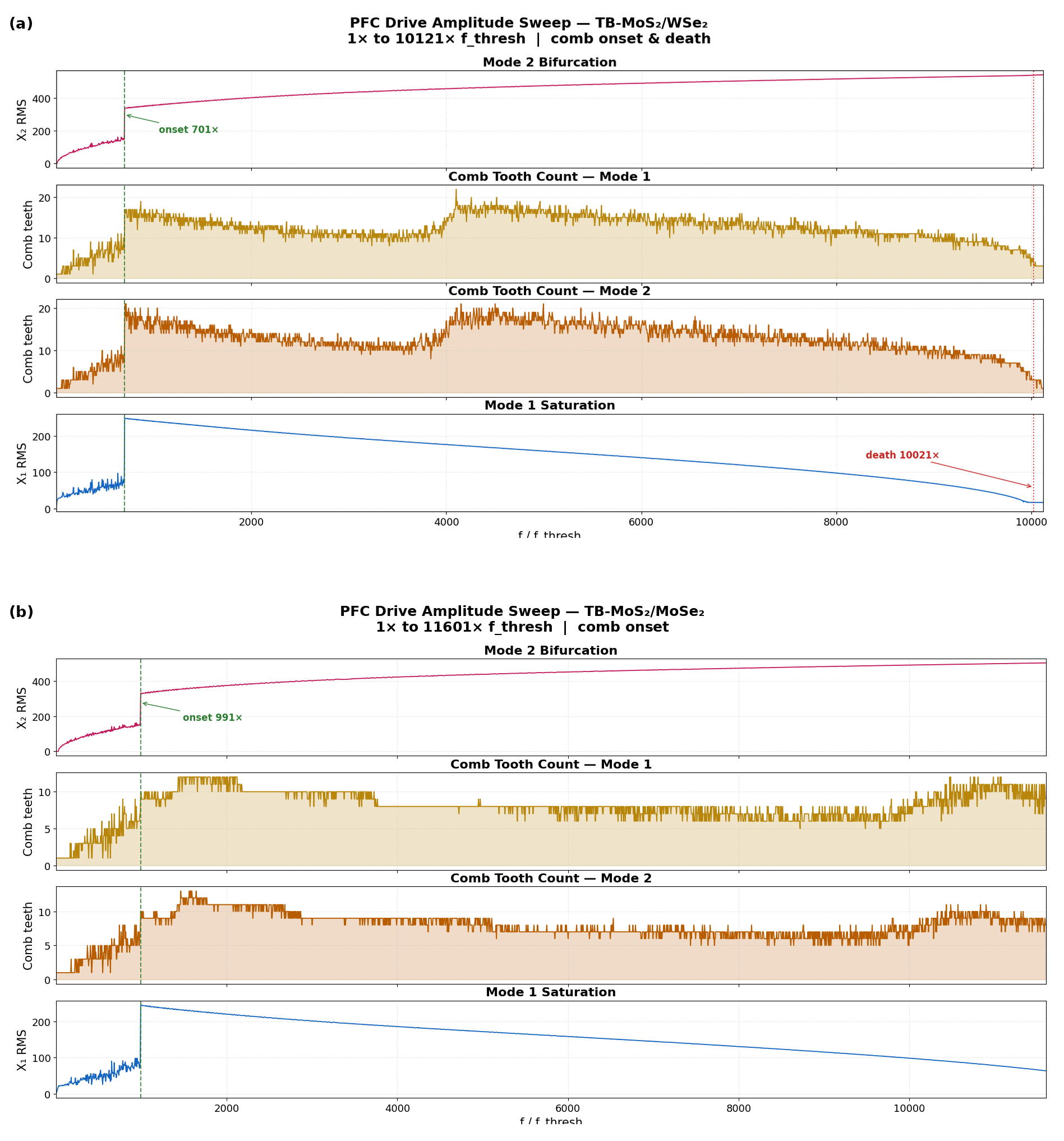}
  \caption{Drive sweeps of the two heterostructures, with drive in units of $\fth$ for each
           compound. (a) \MoSSe{}; (b) \MoSMoSe{}. In each, from top: root-mean-square amplitude
           of the parametrically excited mode ($X_2$), comb tooth count in mode 1 and in
           mode 2, and root-mean-square amplitude of the driven mode ($X_1$). The dashed line marks comb onset and the
           dotted line comb death, which is not reached within the \MoSMoSe{} sweep. All quantities change discontinuously at onset,
           identifying it as a subcritical transition.}
  \label{fig:sweep}
\end{figure*}

Figure~\ref{fig:sweep} resolves the transition. Three quantities are plotted against drive:
the root-mean-square amplitude of the signal mode, the tooth count, and the amplitude of the
driven mode. All three change discontinuously at the same drive. Within an increment of the
sweep, from $696\,\fth$ to $701\,\fth$, the envelope variance rises from $2.1\times10^{3}$
to $1.37\times10^{4}$, the tooth count increases from 7 to 16 in mode 1 and from 7 to 17 in mode 2, and the amplitude of the driven-mode saturates abruptly and then declines as energy is transferred
to the comb. The simultaneous
discontinuity in three independent observables identifies the onset as subcritical: the
trivial solution and a finite-amplitude comb solution coexist over a range of drive, and the
system jumps between them when the trivial branch loses stability. This is consistent with
the negative effective nonlinearity identified by Qi \textit{et al.}~\cite{Qi2020} as the
condition for a hard transition.

Following the jump, the tooth count remains between 8 and 17 across the plateau, with a median of 13 and 
excursions to 22 near $4\times10^{3}\,\fth$, and declines slowly toward death of combs. The
driven-mode amplitude decreases monotonically over the same range, which is the direct
signature of energy being drained from the pump into the comb. Beyond a critical drive, the
comb ceases: at $1.0021\times10^{4}\,\fth$ the tooth count returns toward unity and the
driven-mode amplitude falls to a constant level. The driven mode has saturated to the point
that the modulation depth available to Eq.~(\ref{eq:ode2}) no longer sustains parametric
growth against the damping. Figure~\ref{fig:sweep}(b) shows that \MoSMoSe{} undergoes the same discontinuous jump in tooth count in both modes at its own onset, from 7
to 10 between $986\,\fth$ and $991\,\fth$, and that its comb persists to the end of the
sweep at $1.16\times10^{4}\,\fth$ without collapse.

\subsection{Complexity and the state preceding the comb}
\label{sec:entropy}

\begin{figure*}
  \centering
  \includegraphics[width=\linewidth]{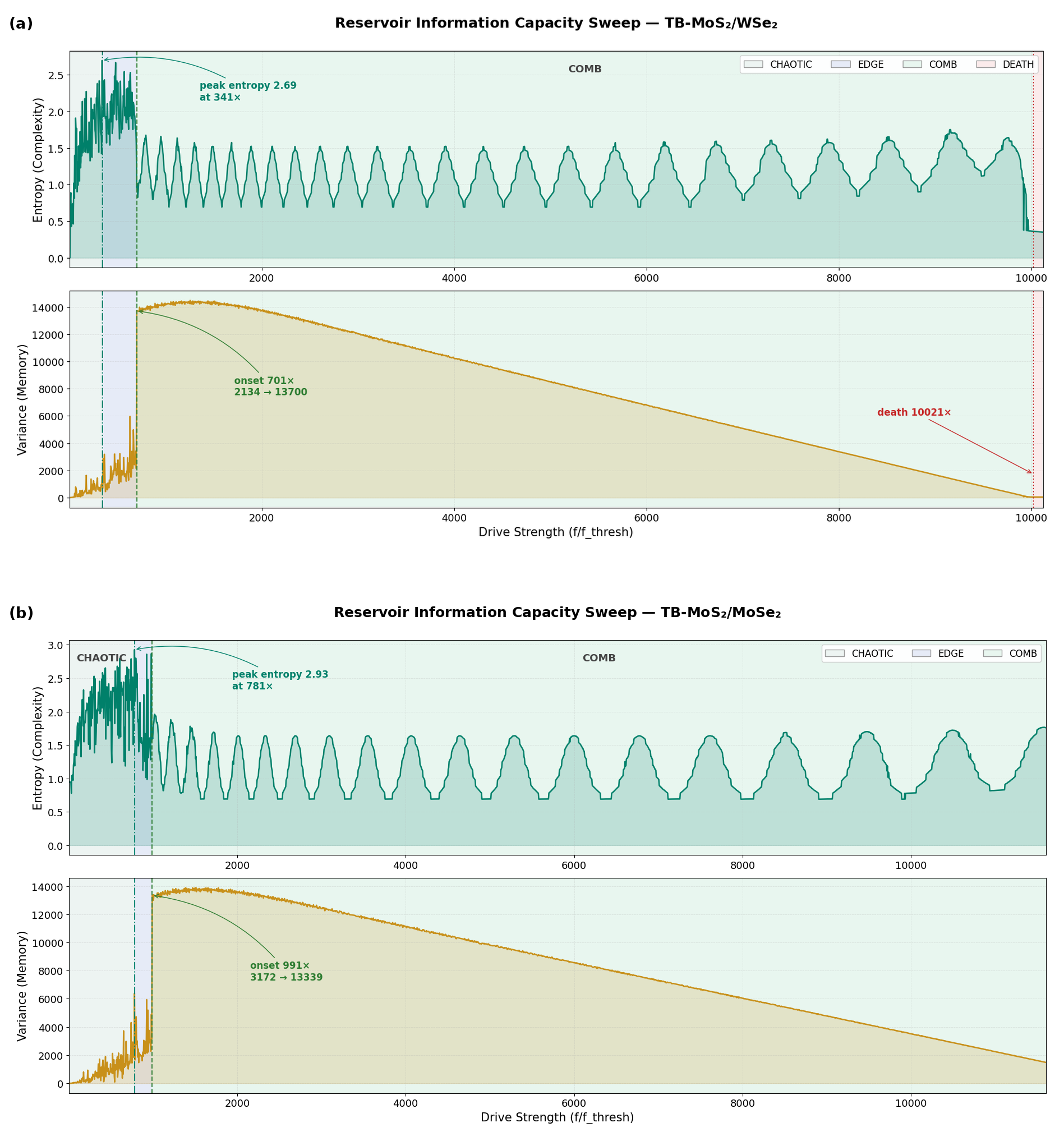}
  \caption{Spectral entropy (upper trace of each panel) and envelope variance of the driven
           mode (lower trace) against drive. (a) \MoSSe{}; (b) \MoSMoSe{}.
           Shaded bands mark the detected regimes; the dash-dotted line marks the entropy
           maximum, the dashed line comb onset and the dotted line comb death (\MoSSe{} only). In both
           compounds the entropy maximum precedes the onset, at which the entropy falls and
           the variance jumps to its plateau.}
  \label{fig:entropy}
\end{figure*}

Figure~\ref{fig:entropy} plots the spectral entropy and the envelope variance across the
sweep. In \MoSSe{}, below $\sim 3.4\times10^{2}\,\fth$, the entropy climbs irregularly, the
variance remains small, and the spectrum is broad and unstructured. The spectral entropy peaks at $H=2.6941$ at $341,\fth$, while the envelope variance remains at $1.55\times10^{3}$, well below its eventual plateau. The comb forms only later, at $701,\fth$, accompanied by a decrease in entropy to between $0.36$ and $1.75$ and a sharp rise in variance to a plateau near $1.4\times10^{4}$. This sequence distinguishes the initial growth of spectral complexity from the subsequent formation of a regular comb. Above $1.0021\times10^{4},\fth$, the variance collapses and the comb disappears.

The entropy maximum therefore precedes comb formation rather than coinciding with it, and the two diagnostics exhibit an anticorrelated response across the transition: the comb is characterised by large-amplitude modulation and low spectral complexity, whereas the most complex state occurs in the disordered regime immediately below onset. The same ordering is observed in \MoSMoSe{}. Its entropy peaks at $2.9287$ at $781,\fth$, where the variance is $6.4\times10^{3}$, before the comb emerges at $991,\fth$. Above onset, the entropy decreases and remains between $0.69$ and $1.96$, consistent with the transition from a spectrally disordered state to a more regular comb.

These findings have direct implications for using the system as a physical reservoir. Effective reservoir computing requires both a rich transformation of the input, reflected here in the spectral entropy, and sufficient dynamical memory, for which the envelope variance serves as a proxy. These properties peak in different operating regimes. The comb plateau offers strong amplitude modulation but relatively low spectral complexity, whereas the low-drive regime provides neither property reliably. The most promising operating window therefore lies between the entropy maximum and comb onset, where spectral complexity remains high while the variance is already increasing: approximately $341$–$701,\fth$ in \MoSSe{} and $781$–$991,\fth$ in \MoSMoSe{}. This window lies below the comb-forming transition, contrary to what the comb spectra alone might suggest, but consistent with the broader observation that recurrent networks can perform best near the boundary between ordered and chaotic dynamics~\cite{Bertschinger2004}. These operating windows are candidate regimes for reservoir computing; their computational utility would need to be established through direct benchmarking of memory capacity and task performance.

\section{Discussion}
\label{sec:discussion}

\subsection{What the twist and chemistry supplies}

The two compounds \MoSSe{} and \MoSMoSe{}  allow the effects of moiré geometry to be distinguished from those of chemical composition, since they share the same twist angle and one constituent layer. In both systems, the resonance condition is readily satisfied: more than 40 mode pairs lie within $0.05$~THz of $\omega_1 = 2\omega_2$, and the best pair in each compound deviates from exact resonance by less than $0.8\%$. This abundance of near-resonant pairs arises from the sevenfold increase in the number of phonon branches due to \rootseven{} zone folding and is largely insensitive to the chemical identity of the second layer.

The nature of the comb differs between the two compounds. Table~\ref{tab:compare} makes the comparison. When measured relative to their respective thresholds, \MoSSe{} exhibits an earlier comb onset at $701,\fth$, compared with $991,\fth$ for \MoSMoSe{}, and supports more comb teeth, reaching 22 versus 12. Conversely, \MoSMoSe{} attains the higher peak spectral entropy ($2.9287$ versus $2.6941$), and its comb persists to the end of the drive sweep. These differences are too pronounced to be explained by the modest factor of $1.20$ in $\kt$ alone. Instead, they are consistent with the quality-factor hierarchy established in Sec.~\ref{sec:pairs}: the signal-mode damping, which differs by a factor of $5.7$, governs how many successive cascade steps can be sustained before the resulting sidebands fall below the detection threshold.

\begin{table}[tbp]
  \caption{Comparison of the phononic frequency comb characteristics of the two heterostructures.}
  \label{tab:compare}
  \begin{ruledtabular}
  \setlength{\tabcolsep}{4pt}
  \begin{tabular}{lcc}
    Quantity & \MoSSe{} & \MoSMoSe{} \\
    \colrule
    Resonant pairs found          & 60    & 48    \\
    Detuning (\%)                 & 0.77  & 0.57  \\
    $\lvert\kt\rvert$ (dimensionless) & $1.00\times10^{-4}$ & $8.37\times10^{-5}$ \\
    Comb onset ($\fth$)           & 701   & 991   \\
    Teeth at onset (mode 1)       & 7 $\to$ 16 & 7 $\to$ 10 \\
    Teeth at onset (mode 2)       & 7 $\to$ 17 & 7 $\to$ 10 \\
    Maximum teeth (mode 1 / 2)    & 22 / 21 & 12 / 13 \\
    Tooth spacing (GHz)           & $\approx 77$ & $\approx 107$ \\
    Peak entropy                  & 2.6941  & 2.9287  \\
    Drive at peak entropy ($\fth$)& 341   & 781   \\
    Plateau variance (max.)       & $1.45\times10^{4}$ & $1.39\times10^{4}$ \\
    Comb death ($\fth$)           & $1.0021\times10^{4}$ & $>1.16\times10^{4}$ \\
  \end{tabular}
  \end{ruledtabular}
\end{table}

Why the coupling should be somewhat larger in \MoSSe{} follows from what $\kt$ measures. Equation~(\ref{eq:V3})
weights the third-order force constants by the product of three eigenvector components, so it is
large when the two modes have substantial amplitude on the same atoms and the anharmonicity there
is strong. Replacing Mo by W and S by Se changes both the mass and the bonding of the second layer,
so the modes of the two layers differ in frequency and the ones that do hybridise do so over a
region where the interlayer potential is strongly anharmonic. Replacing only S by Se leaves the
layers almost nearly identical, their modes  nearly degenerate, and the hybridised eigenvectors
more evenly distributed, which reduces the overlap integral. The resulting design principle is to maximise the chemical contrast between the layers, consistent with the enhanced anharmonic decay observed in Raman measurements of twisted \MoSt{}~\cite{Alencar2025}.

A further symmetry consideration is that the moiré cells belong to point group $3$ ($C_3$) and lack an inversion centre. In a centrosymmetric structure, $\kss$ would be nonzero only for driven modes of even parity. The absence of inversion symmetry lifts this selection rule, allowing otherwise symmetry-forbidden cubic couplings and making all near-resonant pairs potentially eligible. This favours non-centrosymmetric commensurate twist angles for phononic frequency comb generation.

\subsection{Limitations}
\label{sec:limits}

The quality factors are the least certain quantities reported here. With only ten random configurations, the third-order force constants are weakly constrained. Since the scattering rate in Eq.~(\ref{eq:gamma}) depends on $\lvert V^{(3)}\rvert^{2}$, an underdetermined fit may underestimate $\Gamma$ and consequently overestimate (Q). The threshold $\fth$ scales as $1/(Q_1Q_2)$; thus, overestimated quality factors lead to an artificially low absolute threshold. However, the drive sweep is expressed in units of $\fth$, eliminating the overall threshold scale from the comparison. The sequence of transitions within this normalized sweep—including the entropy maximum preceding comb onset, the discontinuity at onset, the plateau, and the eventual extinction—depends on (Q) through the damping relative to the detuning. Larger damping would therefore shift the drive range over which the comb persists. The comparison between compounds is less sensitive to this uncertainty because both are treated using the same computational protocol. Increasing the number of random configurations is the most direct way to improve the reliability of the quality factors.

The coupling constants are better determined than the linewidths, as they are single
tensor elements contracted with eigenvectors rather than a sum over the whole Brillouin zone. Nevertheless, they inherit the same regularisation bias and should be interpreted as order-of-magnitude estimates rather than quantities known to a precisely specified accuracy.

Spin–orbit coupling (SOC) was neglected. This approximation is well justified for $MoS_2/MoSe_2$, but is less certain for $MoS_2/WSe_2$, where the tungsten-containing layer may introduce percent-level shifts in optical phonon frequencies. Because comb formation depends on the frequency ratio $\omega_1/\omega_2$, a common-mode frequency shift cancels, leaving only differential shifts relevant to the resonance condition. We expect the resulting change in $\omega_1/\omega_2$ to be smaller than the existing detuning of (0.77$\%$), although this has not been verified. Including SOC is therefore a priority for future calculations, particularly for heterostructures containing heavier elements.

Finally, the dynamics are described by a two-mode model. The 42-atom unit cell supports 126 phonon modes, and the scan in Sec.~\ref{sec:pairs} identifies more than forty near-resonant mode pairs in each compound. Whether these additional pairs remain spectators over the drive range considered or support competing combs lies beyond the scope of the two-mode reduction. The tooth-count fluctuations observed across the plateau in Fig.~\ref{fig:sweep} are consistent with a two-mode system sampling nearby states, but could also indicate weak coupling to a third mode.

\FloatBarrier
\section{Conclusions}

We have computed phononic frequency combs in twisted bilayer transition metal dichalcogenide
heterostructures from density functional theory, obtaining the mode frequencies, quality
factors and anharmonic coupling constants from the harmonic and third-order force constants from first-principles. Both \MoSSe{} and \MoSMoSe{} at $21.787^{\circ}$ generate combs. In \MoSSe{} the tooth count, the
signal-mode variance and the driven-mode amplitude all change abruptly within a single sweep
increment at $701\,\fth$, which identifies the onset as subcritical rather than supercritical.

The spectral entropy peaks immediately before comb formation. For reservoir computing, this suggests that the most productive operating regime lies just below the comb threshold, where the spectral complexity remains high while the oscillation amplitude is already increasing, rather than within the comb state itself. 

The two compounds exhibit comparable coupling strengths, with $\kt$ differing by a factor of 1.20, yet their comb dynamics differ substantially. \MoSSe{} exhibits a lower comb excitation threshold and sustains a larger number of teeth, whereas \MoSMoSe{} achieves a higher peak spectral entropy and maintains its comb even for larger drives. 

Any commensurate, non-centrosymmetric twist in a van der Waals bilayer can be screened using this approach before device fabrication. The screening criteria comprise two distinct ingredients: the twist geometry determines the availability of near-resonant modes, while their lifetimes govern the drive required to exploit that resonance.

\FloatBarrier
\begin{acknowledgments}
The authors would like to thank the Department of Electrical and Electronics Engineering at BITS Pilani
Dubai Campus for computational resources and support.
\end{acknowledgments}
\vspace{1.75cm}


\end{document}